\documentclass[prl,twocolumn,showpacs,superscriptaddress,nofootinbib]{revtex4-1}
\usepackage{dcolumn}
\usepackage{bm}
\usepackage{graphicx}
\usepackage{amsfonts}
\usepackage{bbm}
\usepackage{dsfont}
\usepackage{float}
\usepackage{graphicx}
\usepackage{bm}
\usepackage{amssymb}
\usepackage{multibib}
\usepackage{color}
\usepackage{enumerate}
\usepackage{amsmath}
\usepackage{amstext}
\usepackage{latexsym}
\usepackage{braket}
\usepackage{appendix}
\usepackage{textcomp}
\usepackage[usenames,dvipsnames]{xcolor}
\usepackage[colorlinks=true,citecolor=Blue,linkcolor=RubineRed,urlcolor=Blue]{hyperref}

\def\la{\langle}
\def\ra{\rangle}

\def\tr{{\rm Tr}}

\newcommand{\beq}{\begin{equation}}
\newcommand{\eeq}{\end{equation}}
\newcommand{\beqa}{\begin{eqnarray}}
\newcommand{\eeqa}{\end{eqnarray}}

\begin{document}
\title{Spectral Filtering Induced by  Non-Hermitian Evolution with \\ 
Balanced Gain and Loss:  Enhancing  Quantum Chaos}

\author{Julien Cornelius}
\affiliation{Department  of  Physics  and  Materials  Science,  University  of  Luxembourg,  L-1511  Luxembourg, Luxembourg}
\author{Zhenyu Xu}
\affiliation{School of Physical Science and Technology, Soochow University, Suzhou 215006, China}
\author{Avadh Saxena}
\affiliation{Theoretical Division and Center for Nonlinear Studies, Los Alamos National
Laboratory, Los Alamos, NM 87545, USA}
\author{Aur\'elia Chenu}
\affiliation{Department  of  Physics  and  Materials  Science,  University  of  Luxembourg,  L-1511  Luxembourg,  Luxembourg}
\author{Adolfo del Campo}
\affiliation{Department  of  Physics  and  Materials  Science,  University  of  Luxembourg,  L-1511  Luxembourg,  Luxembourg}

\begin{abstract}
The dynamical signatures of quantum chaos in an isolated system are captured by the spectral form factor, which exhibits as a function of time a dip, a ramp, and a plateau, with the ramp being governed by the correlations in the level spacing distribution. While decoherence generally suppresses these dynamical signatures, the nonlinear non-Hermitian evolution with balanced gain and loss (BGL) in an energy-dephasing scenario can enhance manifestations of quantum chaos.
In the Sachdev-Ye-Kitaev model and random matrix Hamiltonians, BGL  increases the span of the ramp, lowering the dip as well as the value of the plateau, providing an experimentally realizable physical mechanism for spectral filtering. The chaos enhancement due to BGL is optimal over a family of filter functions that can be engineered with fluctuating Hamiltonians.
\end{abstract}

\maketitle

Non-Hermitian physics offers an exciting arena at the frontiers of physics where nonequilibrium phenomena govern.
In such a scenario, the evolution no longer conserves energy and is characterized by dissipation \cite{Ashida20}.
Its relevance was soon appreciated in nuclear theory \cite{Gamow28}, chemical dynamics \cite{Levine11} and quantum optics \cite{Plenio98}, but its manifestations span over a wide diversity of fields such as mechanics, photonics and active matter \cite{Ashida20}. Condensed matter theory of many-body physics is actively being extended in this setting.
An exciting frontier focuses on the interplay between the nonequilibrium dynamics of non-Hermitian systems and quantum chaos.

Statistical features of the energy spectrum of an isolated system play a crucial role in the dynamics and differentiate systems exhibiting quantum chaos from others governed by integrability, many-body localization, etc. The level spacing distribution varies in these systems from the Wigner-Dyson distribution to an exponential decay \cite{Guhr98,Haake}, although a clearcut classification between chaotic and integrable systems is often more subtle \cite{Borgonovi16}.
Quantum chaos is associated with correlations among energy levels, as revealed by the two-point energy-level distribution \cite{Bohigas84}. In particular, correlations between energy levels can be conveniently captured by the spectral form factor (SFF) defined in terms of the Fourier transform of the energy spectrum \cite{Leviandier86,WilkieBrumer91,Alhassid93,Ma95} or its complex Fourier transform, which can be written in terms of the partition function of the system analytically continued to complex temperature
\cite{Cotler2017,Dyer2017,delcampo17}. Spectral correlations also manifest directly in the Loschmidt echo \cite{Gorin06,Jacquod09,Yan20} and the quantum work statistics \cite{Chenu17work,GarciaMata17,Arrais18,Chenu19}. The identification of universal features in spectral statistics is generally eased by the use of spectral filters which has become ubiquitous in theoretical and numerical studies of quantum systems, chaotic or not \cite{Wall95,Mandelshtam97,Gharibyan18,Schiulaz19,Suntajs20}.

At the time of writing, it remains unclear how signatures of quantum chaos are modified in open quantum systems \cite{Haake,Braun03}. To tackle this question, one can make
 use of random-matrix tools \cite{MethaBook}, as done conventionally for Hamiltonian systems in isolation, but to describe quantum operations instead \cite{Haake,Denisov19,Can19,Can19b,Sa19,Sa20,Wang20}.
These efforts follow the spirit of Hamiltonian quantum chaos in adopting a statistical approach to identify generators of evolution (or quantum channels) compatible with a set of symmetries.
In addition, diagnostic tools to characterize open quantum chaotic systems remain to be developed. Efforts to this end can be split into two groups. The first one focuses on the spectral statistics of the generator of the evolution of an open quantum system, whether it is a Liouvillian governing the rate of change of a quantum state or a quantum channel \cite{Haake92,Sa19,Akemann19,li2021spectral}. This powerful approach leverages the elegance and universality of the Hamiltonian counterpart but seems better suited to capture the dynamics of complex systems in complex environments than to explore how Hamiltonian chaos is altered by decoherence. The second approach uses information-theoretic quantities such as the fidelity or Loschmidt echo and can provide a clear separation between the role of the environment and the spectral features of the system, singling out the correlations in the system's spectral properties that contribute directly to the quantum dynamics \cite{Xu19,delCampo2020,Xu21SFF}.

Across the quantum-to-classical transition, decoherence brings out signatures of classical chaos \cite{Cucchietti03,Zurek03,Habib06}. By contrast,
decoherence generally suppresses dynamical manifestations of quantum chaos stemming from energy level correlations \cite{Rammensee18,Xu19,delCampo2020,Touil21,Xu21SFF,Zanardi21,Anand21}.
In this Letter, we explore the non-Hermitian evolution of chaotic quantum systems. We show that in this setting environmental decoherence can enhance dynamical signatures of quantum chaos. Specifically, we consider the nonlinear evolution of energy-diffusion processes under balanced gain and loss, which is shown to act as a spectral filter. Using the Sachdev-Ye-Kitaev model as a paradigmatic test-bed, we demonstrate the amplification of quantum chaos using a fidelity-based generalization of the spectral form factor to open quantum systems, which is amenable to studies in the laboratory.

{\it Balanced gain and loss from null-measurement conditioning.---}
The Markovian evolution of a quantum system in a quantum state $\rho$ can be described by a master equation of the Lindblad form
$
d_t\rho=-i[H_0,\rho]+\sum_{\alpha}\gamma_{\alpha}\left(K_\alpha\rho K_\alpha^\dag-\frac{1}{2}\{K_\alpha^\dag K_\alpha,\rho\}\right),\nonumber
$
in terms of the positive decay rates $\gamma_\alpha\geq 0$ and the bath operators $K_\alpha$ \cite{Lindblad76}.
For our analysis, we rewrite this evolution as follows \cite{Carmichael09},
\beqa
d_t\rho=-i\left(H_{\rm eff}\rho-\rho H_{\rm eff}^\dag \right)+J(\rho),
\label{QJME}
\eeqa
in terms of the effectively non-Hermitian Hamiltonian given by
$
H_{\mathrm{eff}}=H_{0}-\frac{i}{2}\sum_{\alpha }\gamma _{\alpha }K_{\alpha
}^{\dag }K_{\alpha }$
and the jump term
$J(\rho)=\sum_\alpha\gamma_\alpha K_\alpha\rho K_\alpha^\dag$.
In the absence of quantum jumps, the contribution of the latter term can be ignored, and the dynamics
is exclusively governed by the non-Hermitian Hamiltonian.
The trace preserving evolution for such subensemble of trajectories is given by the nonlinear Schr\"odinger equation for null-measurement conditioning
\beqa
\label{BGLME}
d_t\rho=-i\left(H_{\rm eff}\rho-\rho H_{\rm eff}^\dag \right)+i\,\tr\left[\left(H_{\rm eff}-H_{\rm eff}^{\dag}\right)\rho\right]\rho,\nonumber\\
\eeqa
which also arises in non-Hermitian systems in scenarios characterized by balanced gain and loss (BGL) \cite{BrodyGraefe12}.
Thus, BGL dynamics can be derived as the effective evolution of an ensemble of quantum trajectories conditioned on a measurement record with no quantum jumps, e.g., in a system under continuous monitoring \cite{Carmichael09}.
We note that BGL dynamics also emerges naturally in other experimental settings effectively realizing non-Hermitian Hamiltonians with broken parity-time symmetry, in which eigenvalues come in complex conjugate pairs \cite{Bender98,Bender07}; see e.g., \cite{Schindler11,Peng14,Peng14b,Chang14,Feng14}.

{\it Energy dephasing with and without quantum jumps.---}
 Processes characterized by energy dephasing arise naturally in a variety of scenarios, including random quantum measurements \cite{Gisin84,Korbicz17}, clock errors in timing the evolution of a quantum system \cite{Egusquiza99}, and fluctuations in the system Hamiltonian \cite{Milburn91,Adler03}, such as those invoked by wavefunction collapse models \cite{Percival94,Bassi03,Bassi13}. The evolution of the quantum state is then exactly described by
$d_t\rho=-i[H_0,\rho]-\gamma[H_0,[H_0,\rho]]$, with no restriction on $\gamma$ to the weak-coupling limit \cite{Chenu17,Xu19}. 
This can be recast as the master equation (\ref{QJME}) in terms of a non-Hermitian Hamiltonian
$H_{\mathrm{eff}}=H_0-i\gamma H_0^2$
and the quantum jump term
$
J(\rho)=2\gamma H_0\rho H_0.
$
In the absence of quantum jumps, the trace-preserving evolution is described by the nonlinear master equation (\ref{BGLME}).
Given $H_0=\sum_nE_n|n\ra\la n|$, for a generic initial quantum state $\rho(0)=\sum_{nm}\rho_{nm}(0)|n\ra \la m|$, the exact solution can be found by first solving the linear case, dropping the nonlinear term which simply accounts for the correct normalization, and subsequently including its effect. The time-dependent density matrix reads
\beqa
\label{rhotBGL}
\rho(t)=\frac{\sum_{nm}\rho_{nm}(0)e^{-i(E_n-E_m)t-\gamma t (E_n^2+E_m^2)}}{\sum_{n}\rho_{nn}(0)e^{-2t\gamma E_n^2}}|n\ra \la m|.\nonumber\\
\eeqa
With knowledge of the quantum state during time-evolution, we turn our attention to the interplay among Hamiltonian quantum chaos, energy dephasing, and BGL.
In open quantum systems, different quantities have been proposed to characterize dissipative quantum chaos using spectral properties \cite{Haake,Gorin06,Jacquod09,Xu19,Can19,Xu21SFF}. An analogue of the SFF is given by the fidelity between a coherent Gibbs state
 \beqa
|\psi _{\beta }\rangle =\sum_{n}\frac{e^{-\beta E_{n}/2}}{\sqrt{Z_0(\beta )}}|n\rangle,\quad Z_0(\beta)=\tr[e^{-\beta H_0}],
\eeqa
and its time-evolution \cite{delcampo17,Xu19,delCampo2020,Xu21SFF}. For an arbitrary dynamics described by a quantum channel $\Lambda$, $\rho(t)=\Lambda[\rho(0)]$, the analogue of the SFF reads
$F_{t}=\langle \psi _{\beta }|\rho_t|\psi
_{\beta }\rangle$.
In the limit of unitary dynamics generated by a Hermitian Hamiltonian $H_0$, one recovers the familiar expression \cite{Cotler2017,Dyer2017,delcampo17}
$F_{t}=|Z_0(\beta +it)/Z_0(\beta )|^{2}$.
The result under energy-dephasing has been explored in \cite{Xu19,delCampo2020,Xu21SFF}.
Explicit evaluation using the time-dependent density matrix under BGL  \eqref{rhotBGL} yields the SFF
\beqa
\label{Ftbgl}
F_t=\frac{\left|\sum_{n}e^{-(\beta+it)E_n-\gamma t E_n^2}\right|^2}{Z_0(\beta)\sum_{n}e^{-\beta E_n-2t\gamma E_n^2}}.
\eeqa
This expression corresponds to a single system Hamiltonian and is generally to be averaged over a Hamiltonian ensemble
to reflect eigenvalue correlations, unless the system is self-averaging.
The fidelity-based approach to generalize the SFF is thus naturally suited to account for non-Hermitian quantum systems, including the nonlinear evolution characterized by BGL. With these tools at hand, we proceed to explore the fate of the dynamical signatures of quantum chaos in this setting.

\begin{figure}[t]
\begin{center}
\includegraphics[width=1\linewidth]{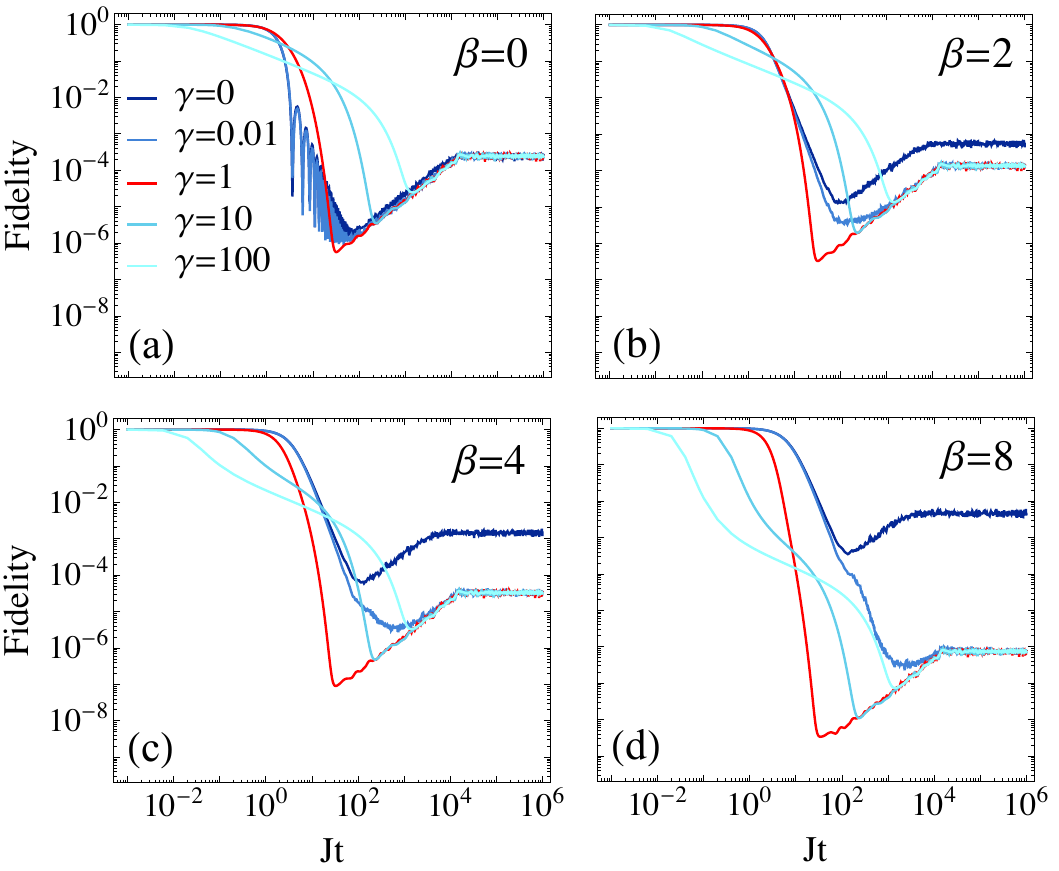}
\end{center}
\caption{Enhancement of quantum chaos under BGL in an energy-dephasing process.
The time-dependence of the fidelity between a coherent Gibbs state and its nonlinear time evolution is shown in the SYK model with $N=26$, after averaging over $100$ realizations of $J_{klmn}$. The enhancement of quantum chaos is more pronounced as the value of $\beta$ is increased.
Under BGL, the dip is enhanced, increasing the span of the ramp associated with fully chaotic dynamics. The
value of the plateau is lowered with respect to the unitary case without a pronounced shift of its onset.
 For larger values of $\gamma$, the features of the SFF in isolation are gradually washed out by the nonunitary evolution. In particular, the span of the ramp is shrunk, without altering the onset of the plateau and its value. } 
\label{Fig1BGL}
\end{figure}

{\it BGL dynamics of the Sachdev-Ye-Kitaev model.---}
For the sake of illustration, we consider the Sachdev-Ye-Kitaev (SYK) model which is known to be maximally chaotic. The Hamiltonian of the SYK model \cite{Sachdev93,AK15}
\begin{equation}
H_0=\frac{1}{4(4!)}\sum_{k,l,m,n=1}^{N}J_{klmn}\chi _{k}\chi _{l}\chi _{m}\chi_{n},
\label{SYK}
\end{equation}%
involves $N$ Majorana fermions satisfying $\{\chi _{k},\chi _{l}\}=\delta _{kl}$ subject to all-to-all random
quartic interactions. The coupling tensor $J_{klmn}$ is completely anti-symmetric, and
independently sampled from a Gaussian distribution
$J_{klmn}\in \mathcal{N}\left( 0,\frac{3!}{(N)^{3}}J^{2}\right)$,
where $J^{2}=\frac{1}{3!}\sum_{lmn}\left\langle J_{klmn}^{2}\right\rangle$
and we set $J=1$ for convenience. Its experimental simulation is the subject of ongoing studies \cite{Danshita17,Laura17,Pikulin17,Babbush19,Luo19} and the features of the SFF in isolation have been characterized in depth \cite{Cotler2017}. It exhibits a decay from unit value towards a correlation hole or dip.
This decay is governed by the density of states and as such, it is not universal. It stops at a characteristic dip time $t_d$. After the correlation hole, quantum chaos governs the evolution giving rise to a ramp as a result of the correlations between different energy levels. Such ramp saturates to a plateau at a second characteristic time $t_p$, as shown in Fig. \ref{Fig1BGL} for $\gamma=0$. The occurrence of the ramp during the interval $(t_d,t_p)$ is a clear manifestation of quantum chaos in the dynamics.
Such dynamical signatures of quantum chaos are however suppressed by decoherence. 
Indeed, energy dephasing, that includes quantum jumps, reduces the depth of the correlation hole and delays the beginning of the ramp, while barely affecting the onset of the plateau \cite{Xu21SFF}.

%
In stark contrast, BGL dynamics is shown to enhance the dynamical signatures of quantum chaos. An explicit computation of the SFF for the SYK model is shown in Fig. \ref{Fig1BGL} averaging over different realizations of the disorder. 
 Results in other paradigms of chaos, such as random-matrix Hamiltonians in the Gaussian Orthogonal and Unitary Ensembles, are detailed in \cite{SM}, with an analytical expression for the latter.
The effective non-Hermitian Hamiltonian accelerates the nonuniversal decay from unit value (associated with the disconnected part of the SFF), thus shifting the onset of the dip.
 BGL provides a physical mechanism to implement the kind of spectral filter proposed to suppress nonuniversal effects from the spectral edges in theoretical and numerical studies \cite{Wall95,Mandelshtam97,Gharibyan18,Schiulaz19,Suntajs20}. Such filters provide an analog of apodization in the time domain, suppressing fringes stemming from the sharp edges of the spectrum in the SFF. As seen from the numerator of (\ref{Ftbgl}), the anti-Hermitian part of the effective non-Hermitian Hamiltonian $H_{\mathrm{eff}}=H_0-i\gamma H_0^2$ gives rise to a Gaussian spectral filter $g(E)=\exp(-\gamma t |E|^2)$, with a strength that increases linearly in time  and width that decreases as $1/\sqrt{\gamma t}$, while the BGL dynamics enhances the signal of the fidelity by making the evolution trace-preserving, giving rise to the denominator in (\ref{Ftbgl}).
The subsequent ramp spans over a stage of the evolution which is not only longer than in the case under energy-dephasing but that also exceeds the ramp interval in the isolated case, e.g., the conventional SFF for unitary dynamics. Away from the infinite temperature limit, the ramp is prolonged  up to two orders of magnitude over the unitary case, see Fig. \ref{Fig1BGL}(d).

%
\begin{figure}[t]
\begin{center}
\includegraphics[width=0.9\linewidth]{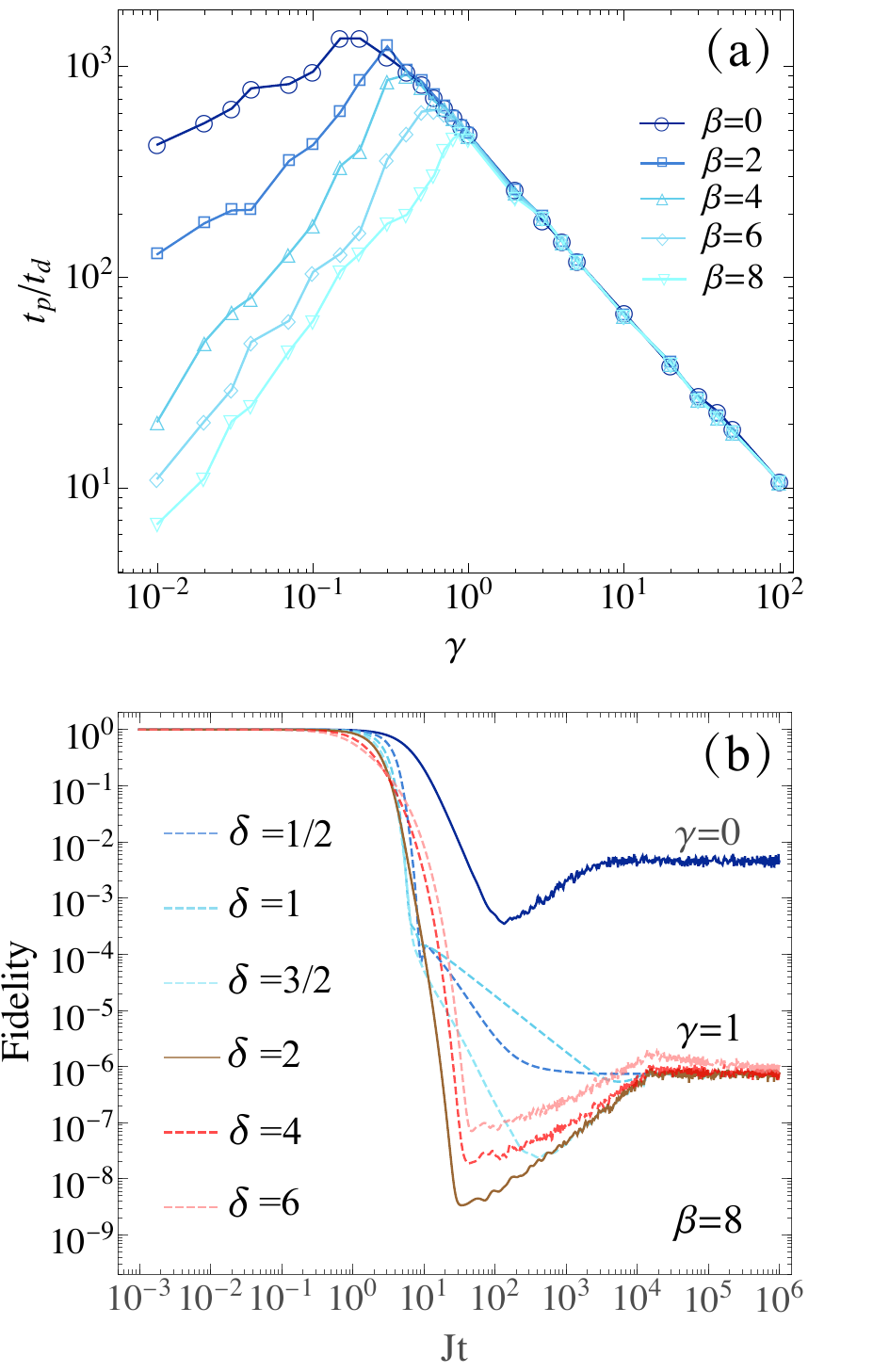}
\end{center}
\caption{(a) Quantifying the enhancement of quantum chaos under BGL in the SYK model. The ratio between the dip and the plateau times is shown as a function of the dephasing rate $\gamma$ for different values of the inverse temperature $\beta$. (b) The Gaussian filter $\delta=2$ that emerges under energy diffusion with BGL is shown to be optimal over the family of filter functions $g(E)=\exp(-|E|^\delta)$ in maximizing the duration of the ramp. Data for $N=26$ is averaged over 100 realizations of $J_{klmn}$.}
\label{Fig2BGL}
\end{figure}
For an isolated system ($\gamma=0$), denoting the degeneracy of an energy level $E_n$ by $N_n$, the asymptotic value of the SFF is given by
$F_p=\frac{1}{Z_0(\beta)^2}\sum_nN_ne^{-2\beta E_n}\geq
 \frac{Z_0(2\beta)}{Z_0(\beta)^2}$ where the lower-bound holds in the absence of degeneracies ($N_n=1$), expected in quantum chaotic systems.
In the infinite-temperature limit, $F_p=1/d$ is set by the inverse of the Hilbert space dimension and thus vanishes with increasing system size.
This asymptotic value is preserved under energy-dephasing, as the long-time quantum state is given by the thermal state $\rho_p=\sum_n\exp(-\beta E_n)|n\ra\la n|/Z_0(\beta)$. However, Fig. \ref{Fig1BGL} shows that the asymptotic value varies in the BGL case.
For $\gamma>0$, the long-time limit of the fidelity reads 
\beqa
\label{longFtbgl}
F_p&\sim& \frac{\sum_{n}N_n^2e^{-2\beta E_n-2\gamma t E_n^2}}{Z_0(\beta)\sum_{n}N_ne^{-\beta E_n-2\gamma t E_n^2}}
\geq \frac{1}{Z_0(\beta)},
\eeqa
where the inequality is saturated for systems lacking degeneracies (with $N_n=1$ $\forall n$), e.g., exhibiting quantum chaos. The discrepancy between the unitary and BGL plateau values is thus enhanced with decreasing temperature.

Even if the plateau value $F_p$ of the SFF varies under BGL, the characteristic time $t_p$ at which this asymptotic value is reached is only weakly affected by BGL with respect to the unitary case and the enhancement of the ramp can be traced to the shortening of the dip time $t_d$ induced by the BGL spectral filter. As the ramp is governed by the eigenvalue correlations stemming from quantum chaos, their enhancement can be quantified by the ratio $t_p/t_d$ as a function of the dephasing rate $\gamma$ that enters the Gaussian filter term in Eq. (\ref{Ftbgl}); see  Fig. \ref{Fig2BGL}(a). There is a critical value of $\gamma$ above which BGL  minimizes the ramp as decoherence mechanisms generally do. However, for values of $\gamma$ below the critical one, the duration of the ramp is enhanced  with increasing dephasing. The enhancement is more pronounced for larger values of $\beta$ for which it exceeds two orders of magnitude.

Given an energy spectrum $\{E_n\}$, obtained by theoretical or experimental means, one may wonder whether other filter functions provide an advantage over the Gaussian filter in enhancing signatures of quantum chaos. 
A filter function $g(E)\geq 0$ yields the modified SFF
 \beqa
 F_t=\frac{\left|\sum_{n}e^{-(\beta+it)E_n}g(E_n)\right|^2}{Z_0(\beta)\sum_{n}e^{-\beta E_n}g(E_n)^2}.
 \eeqa
 We consider the family of filter functions $g(E)=\exp(-\gamma t |E|^\delta)$, which includes the Gaussian case for $\delta=2$.
Those with $\delta\geq 2$ can be engineered by generalized energy dephasing processes conditioned on BGL as discussed in \cite{SM}. For completeness, we also consider $0<\delta<2$.
  Figure~\ref{Fig2BGL}(b) shows that the Gaussian filter is optimal  in the sense that it maximizes the duration of the ramp with respect to the family of higher-order Gaussian filter functions $\exp(-\gamma t |E|^\delta)$ with $\delta\geq0$. More general filters are discussed in \cite{SM}.

{\it Quantum simulation of energy dephasing under BGL.---}
The features presented here are not exclusive to the SYK model and we have reproduced them in other quantum chaotic systems, e.g., random-matrix Hamiltonians, as shown in \cite{SM}. However, the phenomenology described thus far is specific to energy dephasing processes governed by BGL. 
Other open quantum systems unrelated to energy dephasing do not exhibit an enhancement of the dynamical signatures of quantum chaos in the presence of BGL. Indeed, when the quantum jump operators $K_\alpha$ do not commute with the system Hamiltonian $H_0$, the SFF generally loses the signatures of quantum chaos, with or without quantum jumps. Said differently, a general Markovian dehasing evolution (e.g., of the kind considered in \cite{Kolovsky19}) suppresses completely the dip and ramp in the SFF as shown in \cite{SM}. Thus, energy-dephasing stands out as the {\it only} kind of time evolution that can be used to enhance dynamical manifestations of chaos in the laboratory, when conditioned to BGL.
From an experimental point of view, energy dephasing is amenable to quantum simulation by coarse-graining in time the evolution of an isolated system \cite{Xu19,delCampo2020,Xu21SFF}.
The SFF under BGL can be expressed as
\beqa
F_t=\frac{\left|\int_{-\infty}^\infty \mathrm{d}s K(t,s)Z_0(\beta+is)\right|^2}{Z_0(\beta)\int_{-\infty}^\infty \mathrm{d}s \mathrm{d}s ' K(t,s)K(t,s')Z_0[\beta+i(s-s')]}.\nonumber\\
\eeqa
in terms of the kernel $ K(t,s)=\frac{1}{\sqrt{4\pi \gamma t}}e^{-\frac{(t-s)^2}{4\gamma t}}$. Knowledge of the analytically continued partition function $Z_0(\beta+is)$ thus suffices to determine the SFF under BGL. A variety of experimental techniques have been demonstrated to measure the partition function in the complex plane, given its manifold applications that range from the study of Lee-Yang zeroes in critical systems \cite{YangLee52,LeeYang52,Wei12,Wei14} to the full counting statistics of many-body observables \cite{Xu19FCS} and positive operator-valued measures such as work in quantum thermodynamics \cite{Dorner13,Mazzola13}.
A ubiquitous approach relies on single-qubit interferometry that utilizes a two-level system or qubit as a probe \cite{Recati05,Goold11,Knap12,Hangleiter15,ElliottJohnson16}.
Experimental demonstrations include, e.g., NMR systems \cite{Peng15} and ultracold gases \cite{Roncaglia14}.
An alternative experimental approach can be conceived by engineering energy dephasing using noise as a resource \cite{Chenu17,Smith18} and measuring the overlap between the initial coherent Gibbs state and its time-evolution either via learning quantum algorithms \cite{Cincio_2018} or interferometry \cite{ness2021observing}.

In summary, we have considered the nonlinear non-Hermitian evolution of a quantum chaotic system under balanced gain and loss (BGL).
Using a fidelity-based generalization of the spectral form factor we have shown that the interplay between energy-dephasing and BGL enhances the dynamical signatures of quantum chaos by providing an experimentally realizable physical mechanism for spectral filtering, i.e., the optimal filter function of Gaussian type. Spectral filtering has become a ubiquitous tool in theoretical and numerical studies of many-body systems, chaotic or not. As a result, our findings motivate the use of BGL as a generic practical tool to probe the spectral features in complex quantum systems. In addition, our results advance the understanding of dissipative quantum chaos and could be explored in quantum simulators by making use of established experimental techniques such as noise engineering and single-qubit interferometry.

\textit{Acknowledgements.---} It is a pleasure to acknowledge discussions with
Mar Ferri, Fernando J. G\'omez-Ruiz, and Apollonas S. Matsoukas-Roubeas. AC and AdC acknowledge the hospitality of DIPC during the completion of this work.
ZX is supported by the National Natural Science Foundation of China under Grant No. 12074280. This work was
supported in part by the U.S. Department of Energy.

\bibliography{DecoCFT_Lett}

\newpage \appendix

\clearpage

\widetext
\begin{center}
\textbf{{\large Supplemental Material---Spectral Filtering Induced by  Non-Hermitian Evolution with \\ 
Balanced Gain and Loss:  Enhancing  Quantum Chaos}}
\end{center}

\setcounter{equation}{0} \setcounter{figure}{0} \setcounter{table}{0}
\makeatletter
\renewcommand{\theequation}{S\arabic{equation}} \renewcommand{\thefigure}{S%
\arabic{figure}} \renewcommand{\bibnumfmt}[1]{[#1]} \renewcommand{%
\citenumfont}[1]{#1}

\section{Nonlinear master equation for the normalized density matrix}
We consider the master equation in Eq. (1) of the main text, with no jump operator, $d_t \rho = -i (H_{\rm eff} \rho - \rho H^\dagger_{\rm eff})$. This evolution leads to a density matrix with time-dependent norm, $d_t (\tr\rho) = - i \,\tr(H_{\rm eff} \rho - \rho H^\dagger_{\rm eff})$. In order to obtain the evolution of a normalized density matrix, we look at $\tilde{\rho} = \rho / \tr\rho$. It evolves as 
\begin{equation}
\begin{split}
d_t \tilde{\rho} =& \frac{d_t \rho}{\tr\rho} - \frac{\rho}{\tr\rho} \frac{d_t \tr\rho}{\tr\rho} \\
=& -i \left(H_{\rm eff} \tilde{\rho} - \tilde{\rho} H^\dagger_{\rm eff}\right) + i \tilde{\rho}\, \tr\left(H_{\rm eff} \tilde{\rho} - \tilde{\rho} \, H^\dagger_{\rm eff}\right),
\end{split}
\end{equation}
which is Eq. (2) in the main text with $\tilde{\rho}$ denoted as $\rho$.

\section{Spectral form factor with general filter functions}

The canonical SFF
\beqa
F_t=\frac{\left|\sum_{n}e^{-(\beta+it)E_n}\right|^2}{Z_0(\beta)^2}
\eeqa
can be modified under a filter function $g(E)$ as follows
\beqa
F_t=\frac{\left|\sum_{n}e^{-(\beta+it)E_n}g(E_n)\right|^2}{Z_0(\beta)\sum_{j}e^{-\beta E_j}g(E_j)^2}.
\eeqa

In the main manuscript, we have considered the case of energy dephasing that gives rise to a Gaussian filter function in the SFF, upon imposing null-measurement conditioning.
This is associated with a Lindblad equation with a single operator $K_\alpha=K_\alpha^\dag=\sqrt{2}H_0$. More generally, one can consider a Lindblad operator
$K_\alpha=K_\alpha^\dag=\sqrt{2}w(H_0)$, where $w$ is an arbitrary complex function. The corresponding effective non-Hermitian Hamiltonian is 
\beqa
H_{\mathrm{eff}}=H_{0}-i\gamma w(H_0)^\dag w(H_0).
\eeqa 
Under null-measurement conditioning, the initial state $\rho(0)=\sum_{nm}\rho_{nm}(0)|n\ra \la m|$
evolves into
\beqa
\rho(t)&=&\frac{\sum_{nm}\rho_{nm}(0)e^{-i(E_n-E_m)t-\gamma t (|w(E_n)|^2+|w(E_m)|^2)}}{\sum_{n}\rho_{nn}(0)e^{-2t\gamma |w(E_n)|^2}}|n\ra \la m|.
\eeqa
Choosing the coherent Gibbs state as the initial quantum state, the fidelity provides an analog of the SFF under BGL and reads
\beqa
F_t=\frac{\left|\sum_{n}e^{-(\beta+it)E_n-\gamma t |w(E_n)|^2}\right|^2}{Z_0(\beta)\sum_{j}e^{-\beta E_j-2t\gamma |w(E_j)|^2}}.
\eeqa
Naturally, for $w(E_n)=E_n$ one recovers the Gaussian filter, which is optimal in maximizing the duration of the ramp, as shown in Fig. 2 (b) of the main manuscript.

On physical grounds, the general class of energy-dephasing processes can be induced by making use of classical noise as a resource \cite{Chenu17}, that is added to the coupling constant of an operator $w(H_0)$ commuting with the system Hamiltonian $H_0$, i.e, 
\beqa
H_{\rm st}=H_0+\sqrt{2\gamma} \xi(t)w(H_0),
\eeqa
where $ \xi(t)$ denotes a real Gaussian process with zero mean and unit variance and we restrict the function $w$ to be real for simplicity.
The density matrix obtained by averaging over different realizations of the noise is then given by
\beqa
d_t\rho=-i[(H_0),\rho]-\gamma[w(H_0),[w(H_0),\rho]],
\eeqa
which is a Lindblad master equation with a single Lindblad operator $K_\alpha=K_\alpha^\dag=\sqrt{2}w(H_0)$.
Thus, filters other than the Gaussian can be associated with energy-diffusion processes under null-measurement conditioning.

%
\begin{figure}[t]
\begin{center}
\includegraphics[width=1\linewidth]{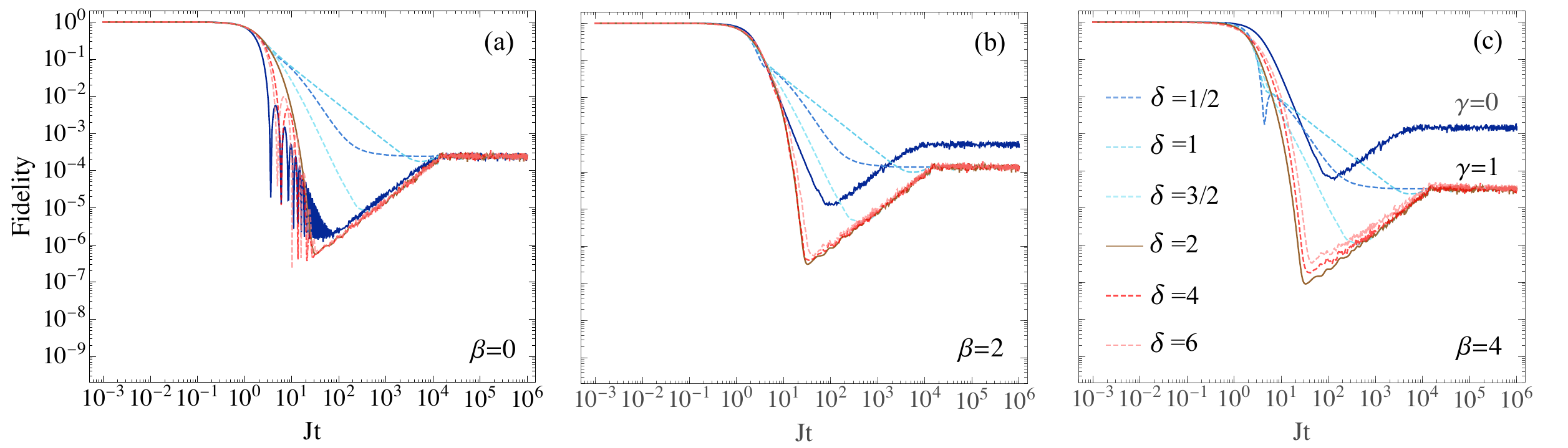}
\end{center}
\caption{SFF in the SYK model as a function of time with higher-order filter functions of the form $\exp(-\gamma t |E|^\delta)$. For $\beta>0$ all filters act similarly on the SFF. The Gaussian filter ($\delta=2)$ is shown to maximize the extension of the ramp as $\beta$ is increased.}
\label{FigBGL_SM1}
\end{figure}
Figure \ref{FigBGL_SM1} displays the behavior of the SFF for the SYK model as a function of time for higher-order Gaussian filters of the form $g(E)=\exp(-\gamma t |E|^\delta)$ (that is, with $|w(E)|^2=|E|^\delta$) for the values $\delta=0,1/2,1,3/2,2,4,6$. The canonical SFF is obtained for $k=0$ (no filter).
 In particular, for any $\beta$, sub-Gaussian filters with $\delta=1/2$ and $\delta=1$ are shown to suppress the dip and ramp of the SFF. For these values of $\delta$, the first derivative of $w(E)$ is nonanalytic near $E=0$ and the filter function acts over all energies, altering even the central part of the spectrum around $E=0$. As a result, such filters distort the SFF of the system under study. 
For $\beta=0$ and $\delta>1$, the actions of hyper-Gaussian filters are effectively indistinguishable from the Gaussian one. As $\beta$ increases,  the duration of the ramp is prolonged for any $\delta\geq2$. However, the maximum extent of the ramp is obtained for $\delta=2$ indicating that the Gaussian filter is optimal.

\begin{figure}[t]
\begin{center}
\includegraphics[width=0.45\linewidth]{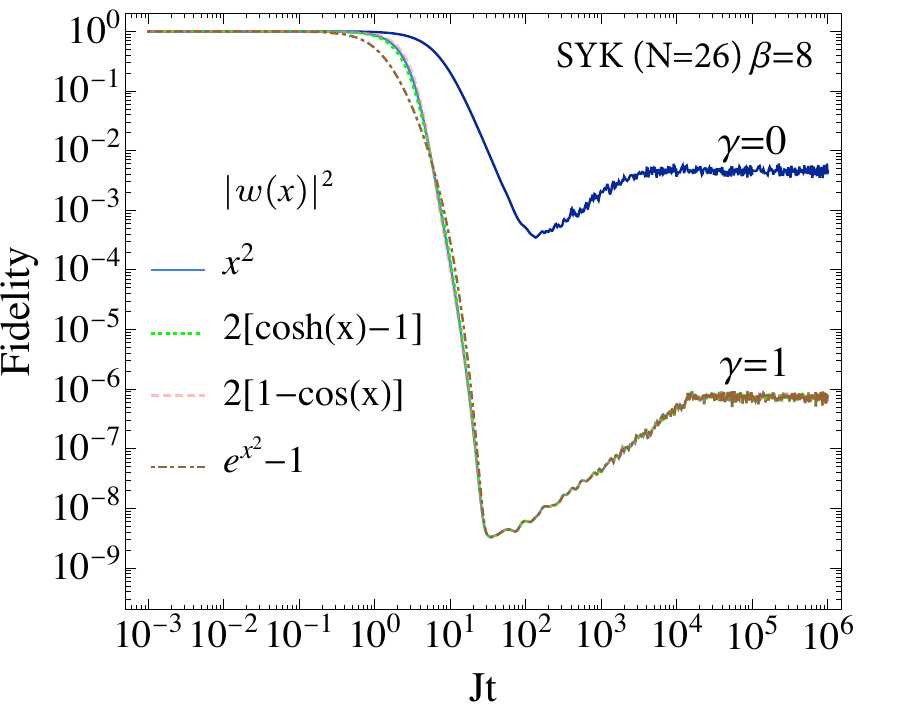}
\end{center}
\caption{SFF in the SYK model as a function of time for filters with a leading Gaussian expansion. Such filters differ only in the extent of the suppression of high-energy eigenstates, that are already filtered. The ramp and the plateau remain invariant.}
\label{FigBGL_SM2}
\end{figure}

To complete the analysis of spectral filtering in the SYK model, we explore the role of other filters admitting a series expansion with a leading Gaussian term. Figure \ref{FigBGL_SM2} shows that their action on the SFF is effectively indistinguishable even for large values of $\beta$. So are the filter functions for small $\gamma t\leq 1$.
%
It is the filtering of eigenvalues satisfying $\gamma tE^2\geq 1$ that plays a crucial role in the SFF. Though this condition varies with time, we note that the main action of the filter is to enhance the nonuniversal decay of the SFF towards the dip, sampling preferentially low-energy states near $E=0$. 

Filters described by more general window functions can be considered and there is abundant literature in the context of filter analysis and apodization on this regard.
While we identify the Gaussian filter as optimal within the family of hyper-Gaussian filters our analysis does not rule out the existence of other filter functions that may be used to enhance signatures in the SFF of spectral correlations associated with quantum chaos. However, such filters are not generally expected to arise naturally in the quantum dynamics as the Gaussian filter does.

\section{BGL dynamics in random-matrix Hamiltonians}
\subsection{Sachdev-Ye-Kitaev (SYK) model}
\begin{figure}[t]
\begin{center}
\includegraphics[width=1\linewidth]{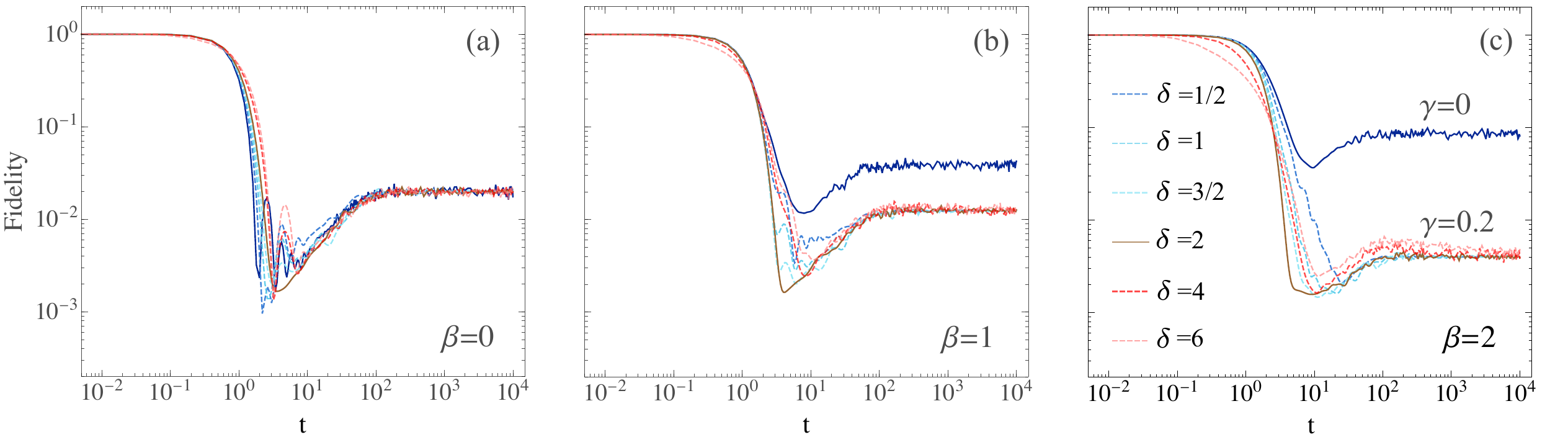}
\end{center}
\caption{SFF in random-matrix Hamiltonians sampled from GOE for higher order Gaussian filter functions. As in the SYK model, the Gaussian filter is optimal and maximizes the extent of the ramp. The data for GOE is obtained with Hilbert space dimension $d=50$, averaging over 200 Hamiltonian instances.}
\label{FigBGL_SM3}
\end{figure}

The results in the main manuscript are illustrated focusing on the SFF of the SYK model as a case study. In this system, the SFF displays distinct regimes encompassing the decay from unit value, dip (correlation hole), ramp, and plateau. These features are common to other chaotic models. To display the universality of energy dephasing conditioned to BGL in maximizing the dynamical manifestations of chaos and prolonging the extent of the ramp, one can consider a different system exhibiting quantum chaos. A paradigmatic setting is that of random-matrix Hamiltonians sampled from the Gaussian Orthogonal Ensemble (GOE), characteristic of systems with time-reversal symmetry. Fig. \ref{FigBGL_SM3} shows the SFF averaged over 200 Hamiltonians $H_0\in{\rm GOE}$ with Hilbert space dimension $d=50$. As in the SYK model, the choice of the Gaussian filter is shown to be optimal for $\beta>0$.
%

\begin{figure}[t]
\begin{center}
\includegraphics[width=1\linewidth]{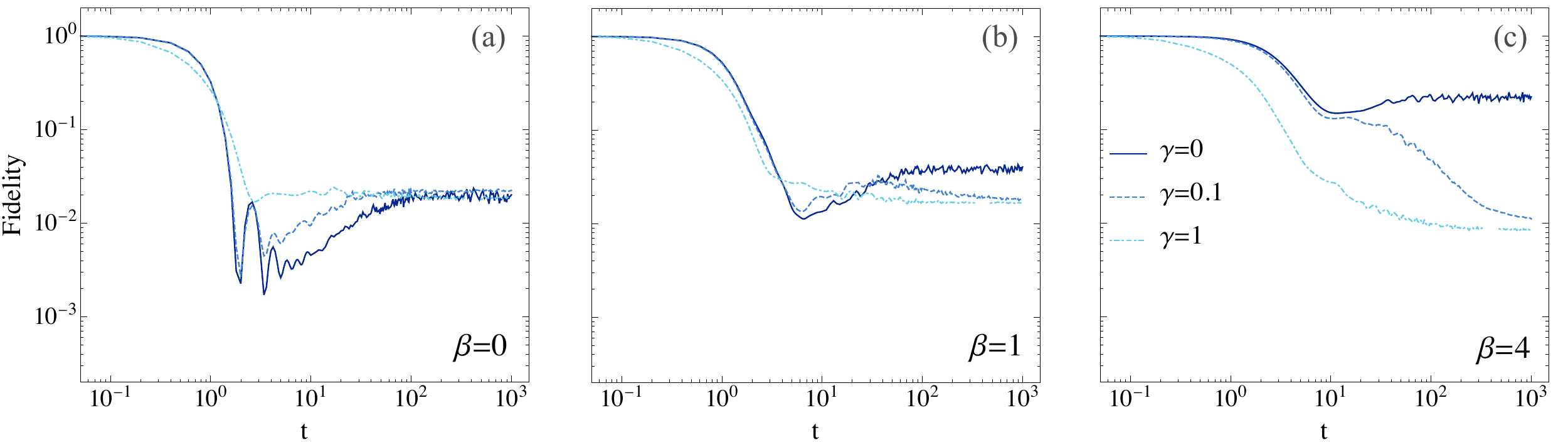}
\end{center}
\caption{SFF in random-matrix Hamiltonians sampled from GOE under BGL associated with generic dephasing, i.e., in the eigenbasis of an operator $X$ that does not commute with the system Hamiltonian $H_0$. In sharp contrast with the BGL associated with energy dephasing, the evolution does not induce spectral filtering and the signatures of chaos in the SFF are suppressed for $\beta\geq 0$ and $\gamma>0$. The data for GOE is obtained with Hilbert space dimension $d=50$, averaging over 100 Hamiltonian instances.}
\label{FigBGL_SM4}
\end{figure}

The situation changes when the Lindblad operators $K_\alpha$ do not commute with the system Hamiltonian. This is illustrated by considering a Markovian master equation $d_t\rho=-i[H_0,\rho]-\gamma[X,[X,\rho]]$ inducing dephasing in the eigenbasis of the observable $K_\alpha=X=X^\dag$, which does not commute with the system Hamiltonian, i.e., $[H_0,X]\neq 0$. This evolution is generated by the effective non-Hermitian Hamiltonian
$H_{\mathrm{eff}}=H_0-i\gamma X^2$ together with the quantum Jump term $J(\rho)=2\gamma X\rho X$. Conditioning the dynamics on the absence of quantum jumps leads to the nonlinear evolution described by Eq. (2) in the main manuscript, which can be numerically integrated, e.g., choosing the initial state as the coherent Gibbs state. This allows computing the SFF shown in Fig. \ref{FigBGL_SM4} for an example in which $H_0$ and $X$ are sampled independently from GOE. For $\gamma=0$ the chaotic features (dip and ramp) of the chaotic system Hamiltonian $H_0$ are displayed in the SFF and are gradually suppressed as $\beta$ is increased from the infinite-temperature limit $\beta=0$. 
For values of $\gamma>0$ and $\beta=0$, the dephasing in the eigenbasis of $X$ gradually washes out the chaotic features, reducing the depth of the dip and flattening the ramp. And $\beta$ is increased, this suppression is further enhanced, leading to a direct decay from unit value towards an asymptotic plateau $F_p$ that depends both on $\beta $ and $\gamma$. This behavior is in sharp contrast with the commuting case shown in Fig. \ref{FigBGL_SM3}, in which BGL leads to spectral filtering maximizing signatures of chaos for $\gamma,\beta>0$.

%

\subsection{Analytical form of the SFF under BGL in the Gaussian Unitary Ensemble GUE($d$)}
The Gaussian Unitary Ensemble (GUE) can be used to describe generic Hamiltonians without imposing time-reversal symmetry and has the advantage of being analytically tractable. Motivated by this observation, we proceed to derive an analytical expression for the ensemble-average SFF in an energy dephasing process characterized by balanced gain and loss.

Starting from the fidelity, Eq. (5) in the main text, we take the ensemble average as 
\begin{equation}\label{eq:ft}
\langle F_t \rangle = \frac{\langle F_{1,t} \rangle }{\langle F_{2,t} \rangle },
\end{equation}
with
\begin{eqnarray}
\langle F_{1,t}\rangle &=&\int dE\left\langle \rho (E)\right\rangle
e^{-2\beta E-2\gamma tE^{2}}+\int dEdE^{\prime }\left\langle \rho (E)\rho
(E^{\prime })\right\rangle e^{-(\beta +it)E-\gamma tE^{2}}e^{-(\beta
-it)E^{\prime }-\gamma tE^{\prime 2}}, \\
\langle F_{2,t}\rangle &=&\int dE\left\langle \rho (E)\right\rangle
e^{-2\beta E-2\gamma tE^{2}}+\int dEdE^{\prime }\left\langle \rho (E)\rho
(E^{\prime })\right\rangle e^{-\beta E-2\gamma tE^{2}}e^{-\beta E^{\prime }}.
\end{eqnarray}%
In performing the average over the GUE with finite dimension $d$, we use the fact that the average spectral
density reads 
\begin{equation}
\langle \rho (E)\rangle =\sum_{n=0}^{d-1}\frac{1}{2^{n}n!\sqrt{\pi }}%
e^{-E^{2}}\mathrm{H}_{n}^{2}(E).
\end{equation}%
In addition, the correlated joint spectral density is given by
\begin{eqnarray}
\langle \rho _{c}^{(2)}(E,E^{\prime }) \rangle&=&\langle \rho (E)\rho (E^{\prime
})\rangle -\langle \rho (E)\rangle \langle \rho (E^{\prime })\rangle \notag
\\
&=&-\sum_{n,m=0}^{d-1}\frac{1}{2^{n}n!\sqrt{\pi }}\frac{1}{2^{m}m!\sqrt{\pi }%
}e^{-E^{2}}\mathrm{H}_{n}(E)\mathrm{H}_{m}(E)e^{-E^{\prime 2}}\mathrm{H}%
_{n}(E^{\prime })\mathrm{H}_{m}(E^{\prime }),
\end{eqnarray}%
where $\mathrm{H}%
_{n}(x)$ denotes the Hermite polynomial of order $n$.

\begin{figure}
\includegraphics[width=1\textwidth]{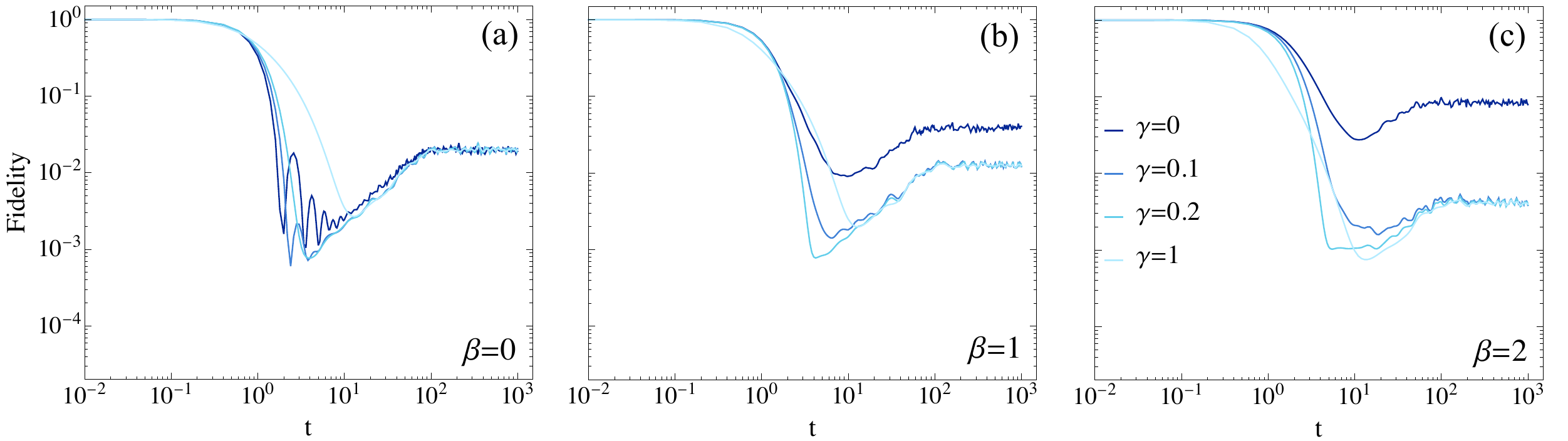}
\caption{SFF in random-matrix Hamiltonians sampled from the GUE ensemble under balanced gain and loss associated with energy dephasing. The SFF is plotted as a function of time for different values of the dephasing strength $\gamma$ and different temperatures. 
The data is shown for a Hilbert space of dimension $d=50$ and obtained from the analytical expression using Eqs. (\ref{eq:ft}), (\ref{eq21}) and (\ref{eq22}). \label{SFFGUE}}
\end{figure}

In order to get an analytical expression of the ensemble-average fidelity, it is useful to define the integrals
\begin{eqnarray}
J_{n,m}^{(\gamma )}(\sigma ) &=&\int_{-\infty }^{\infty }dE \, e^{-\sigma
E-(1+\gamma t)E^{2}}\mathrm{H}_{n}(E)\mathrm{H}_{m}(E), \\
I_{n,m}(\sigma ) &=&\int_{-\infty }^{\infty }dE \,e^{-\sigma E-E^{2}}\mathrm{H}%
_{n}(E)\mathrm{H}_{m}(E),
\end{eqnarray}%
and the time-dependent parameter $\Gamma_t =(1+\gamma t)^{-\frac{1}{2}}$. We use Ref. \cite{Gradshteyn2007a} to find, for $\gamma t \neq 0$,  
\begin{eqnarray}
J_{n,m}^{(\gamma )}(\sigma ) &=&\sqrt{\pi }\Gamma_t e^{\frac{(\Gamma_t \sigma
)^{2}}{4}}\sum_{k=0}^{\min (n,m)}2^{k}k!\binom{m}{k}\binom{n}{k}(1-\Gamma_t
^{2})^{\frac{m+n}{2}-k}\mathrm{H}_{m+n-2k}\left( -\frac{\Gamma_t ^{2}\sigma }{2%
\sqrt{1-\Gamma_t ^{2}}}\right) , \\
I_{n,m}(\sigma ) &=&e^{\frac{\sigma ^{2}}{4}}2^{m}\sqrt{\pi }n!\left( -\frac{%
\sigma }{2}\right) ^{m-n}\mathrm{L}_{n}^{m-n}\left( -\frac{\sigma ^{2}}{2}%
\right) \text{ for }n\leq m,
\end{eqnarray}
where $\mathrm{L}_{n}^{k}(x)$ denotes the generalized Laguerre polynomial.
This allows for an analytic expression of the average functions
\begin{eqnarray} \label{eq21}
\langle F_{1,t}\rangle &=&\sum_{n=0}^{d-1}\frac{1}{2^{n}n!\sqrt{\pi }}%
J_{n,n}^{(2\gamma )}(2\beta )+\left\vert \sum_{n=0}^{d-1}\frac{1}{2^{n}n!%
\sqrt{\pi }}J_{n,n}^{(\gamma )}(\beta +it)\right\vert ^{2}-\sum_{n,m=0}^{d-1}%
\frac{1}{\pi 2^{n}n!{2^{m}m!}}J_{n,m}^{(\gamma )}(\beta +it)J_{n,m}^{(\gamma
)}(\beta -it), \\
\langle F_{2,t}\rangle &=&\sum_{n=0}^{d-1}\frac{1}{2^{n}n!\sqrt{\pi }}%
J_{n,n}^{(2\gamma )}(2\beta )+\sum_{n=0}^{d-1}\frac{1}{2^{n}n!\sqrt{\pi }}%
I_{n,n}(\beta )\sum_{n=0}^{d-1}\frac{1}{2^{n}n!\sqrt{\pi }}J_{n,n}^{(2\gamma
)}(\beta )-\sum_{n,m=0}^{d-1}\frac{1}{\pi 2^{n}n!{2^{m}m!}}I_{n,m}(\beta
)J_{n,m}^{(2\gamma )}(\beta). \: \,  \label{eq22}
\end{eqnarray}
The fidelity readily follows from Eq. \eqref{eq:ft} for $\gamma t \neq 0$, and is illustrated in Fig. \ref{SFFGUE} for $d=50$, in agreement with the findings in the SYK model and random-matrix Hamiltonians in the GOE. In all cases,  BGL in an energy dephasing process enhances quantum chaos for moderate values of $\gamma$ prolonging the extension of the ramp between the dip and the plateau. Additional simulations show that such enhancement can be hidden in systems with a small Hilbert space dimension ($d=10-30$), but that is robust for large values of $d>50$.

\section{Decoherence under BGL: Role of dissipation vs dephasing}
In this manuscript, energy-dephasing processes have been singled out as those allowing for the engineering of spectral filters in the SFF. We note that when energy-dephasing processes are conditioned to BGL in the absence of quantum jumps, the dynamics is governed by dissipation.
Indeed, under BGL the purity of an arbitrary density matrix evolves according to
\beqa
\label{PtEDBGL}
P(t)=\tr[\rho(t)^2]=\frac{\sum_{nm}\rho_{nm}(0)^2e^{-2\gamma t (E_n^2+E_m^2)}}{(\sum_{n}\rho_{nn}(0)e^{-2t\gamma E_n^2})^2}.
\eeqa
This expression has the remarkable feature that whenever the initial state is pure, such that the following factorization holds, $\rho_{nm}(0)=c_n(0)c_m(0)^*$, then $P(t)=1$  for all $ t\geq 0$ and the BGL evolution preserves the purity of the quantum state.
In the fidelity-based definition of the SFF for open quantum systems, the coherent Gibbs state is used as the initial state.
In this case,
\beqa
P(t)=\frac{\sum_{nm}e^{-\beta(E_n+E_m)}e^{-2\gamma t (E_n^2+E_m^2)}}{(\sum_{n}e^{-\beta E_n}e^{-2t\gamma E_n^2})^2}=1,
\eeqa
showing that the state remains pure at all times.
By contrast, the mean energy varies as a function of time according to
\beqa
\la H(t)\ra=\tr[\rho(t)H]=\frac{\sum_{n}\rho_{nm}(0)E_ne^{-2\gamma t E_n^2}}{(\sum_{n}\rho_{nn}(0)e^{-2t\gamma E_n^2})^2}.
\eeqa

\end{document}